\documentstyle[equations,11pt]{article}

\renewcommand{\theequation}{\arabic{equation}}
\begin{document}
\bibliographystyle{plain}
\def\m@th{\mathsurround=0pt}
\mathchardef\bracell="0365 
\def\upbrall{$\m@th\bracell$}
\def\undertilde#1{\mathop{\vtop{\ialign{##\crcr
    $\hfil\displaystyle{#1}\hfil$\crcr
     \noalign
     {\kern1.5pt\nointerlineskip}
     \upbrall\crcr\noalign{\kern1pt
   }}}}\limits}
\def\theequation{\arabic{section}.\arabic{equation}}
\newcommand{\ar}{\alpha}
\newcommand{\aar}{\bar{a}}
\newcommand{\bb}{\beta}
\newcommand{\gm}{\gamma}
\newcommand{\Gm}{\Gamma}
\newcommand{\en}{\epsilon}
\newcommand{\dd}{\delta}
\newcommand{\sg}{\sigma}
\newcommand{\kp}{\kappa}
\newcommand{\ld}{\lambda}
\newcommand{\oa}{\omega}
\newcommand{\be}{\begin{equation}}
\newcommand{\ee}{\end{equation}}
\newcommand{\bea}{\begin{eqnarray}}
\newcommand{\eea}{\end{eqnarray}}
\newcommand{\bse}{\begin{subequations}}
\newcommand{\ese}{\end{subequations}}
\newcommand{\nn}{\nonumber}
\newcommand{\bR}{\bar{R}}
\newcommand{\bP}{\bar{\Phi}}
\newcommand{\bS}{\bar{S}}
\newcommand{\bU}{\bar{U}}
\newcommand{\bW}{\bar{W}}
\newcommand{\vf}{\varphi}
\newcommand{\sn}{{\rm sn}}
\newcommand{\wh}{\widehat}
\newcommand{\wt}{\widetilde}
\newcommand{\ut}{\undertilde}
\newcommand{\ip}{{i^\prime}}
\newcommand{\jp}{{j^\prime}}
\begin{flushright}
solv-int/9603006 \\
21 March 1996 
\end{flushright}
\begin{center} 
{\large{\bf Dynamical $r$-matrix for the elliptic Ruijsenaars-Schneider 
system}}
\vspace{.4cm} 

F.W. Nijhoff  and V.B. Kuznetsov \footnote{
On leave from Department of Mathematical and Computational Physics, 
Institute of Physics, St.Petersburg University, St.Petersburg 198904, 
Russia} \vspace{.2cm} \\ 
{\it Department of Applied Mathematical Studies \\
University of Leeds, Leeds LS2 9JT, U.K.}
\vspace{.2cm}

E.K. Sklyanin \footnote{On leave from Steklov Mathematical Institute,
Fontanka 27, St.Petersburg 191011, Russia}
\vspace{.2cm} \\ 
{\it Research Institute for Mathematical Sciences \\
Kyoto University, Kyoto 606, Japan}
\vspace{.2cm}

O. Ragnisco\vspace{.2cm} \\ 
{\it Dipartimento di Fisica, Terza Universit\`a di Roma \\ 
Via Vasca Navale 84, Roma, Italy} 
\end{center} 
\vspace{.4cm}
\centerline{\bf Abstract}
\vspace{.2cm}

\noindent
The classical $r$-matrix structure for the generic elliptic 
Ruijsenaars-Schneider model is presented. It makes the integrability 
of this model as well as of its discrete-time version 
that was constructed in a recent paper manifest. 
\vskip 1cm

\noindent 
Submitted to J. Phys. A: Math. Gen.
\vskip 2cm
\pagebreak

\section{Introduction} 
\setcounter{equation}{0} 
\setcounter{footnote}{0}

The problem of finding a classical $r$-matrix structure 
for the Calogero-Moser (CM) type of
models aroused some attention a few years ago, 
cf. \cite{AT,Skly,BS}. The fact that this had remained an open problem 
until relatively recently lies probably in the specific feature 
that for these models the $r$-matrix turns out to be of dynamical 
type, i.e. it depends on the dynamical variables. Similar 
features have been found in other integrable many-body problems as well,  
e.g. systems separable in the generalized ellipsoidal coordinates 
\cite{EEKT}. 
 The difficulty presented by the dynamical aspect of the $r$-matrix
is that the Poisson algebra of a model, whose structural
constants are given by a dynamical $r$-matrix is, generally speaking,  
no longer closed, and that there is no closed-form Yang-Baxter equation 
defining the $r$-matrix. So far, only for one particular example ---
the spin generalisation of the Calogero-Moser model --- 
a proper algebraic setting (Gervais-Neveu-Felder equation)
is found \cite{ABB} which also allows to quantize the model. For other
models finding the algebraic interpretation of the dynamical $r$-matrix and,
respectively, solving the quantization problem are still open questions.
Thus, the use of such dynamical 
$r$-matrices so far is very restricted. Nonetheless, the existence 
of such structures is probably significant for the integrability of the 
model --- even though one has not been able thus far to deal very effectively 
with these structures --- and it is foreseeable that they will play a role in 
the further understanding of the models, maybe even on the quantum level. 

One of the most important integrable many-body systems is the 
relativistic variant 
of the Calogero-Moser model, the so-called Ruijsenaars-Schneider (RS) model 
introduced in \cite{Ruijs1,Ruijs2}. 
Its importance lies in the fact that 
it can be considered as a $q$-deformation of the CM model and as such the 
corresponding quantum model is realised in terms of commuting difference 
operators whose eigenfunctions are given in terms of Macdonald polynomials, 
cf. e.g. \cite{Diej,Macd}. 
On the classical level a dynamical $r$-matrix was found 
only very recently in \cite{AR} for the rational and trigonometric 
(hyperbolic) 
cases, although a special parameter-case was already treated in an earlier 
paper,
\cite{BB}. A geometric interpretation was given in a recent preprint, cf. 
\cite{Sur}. 
So far, no results have been found for the full elliptic case. That, 
in fact, is the 
subject of the present 
paper where we will present the dynamical $r$-matrix structure for the 
RS model 
in the generic elliptic case, thus generalising the previous results of 
\cite{AR}--\cite{Sur}. 

\section{Ruijsenaars-Schneider model and its discrete-time version} 
\setcounter{equation}{0} 

The equations of motion of the RS model in its generic (elliptic) form
read
\bse \label{eq:RCM} 
\be \label{eq:RCMa} 
\ddot{q}_i = \sum_{j\ne i} \dot{q}_i \dot{q}_j v(q_i - q_j)\  \ , 
\   \ i=1,\dots,N\  , 
\ee
where the potential $v(x)$ is given by 
\be\label{eq:RCMb} 
v(x) = \frac{\wp^\prime(x)}{\wp(\ld) - \wp(x)}\   , 
\ee \ese 
in which $\wp(x)=\wp(x|\oa_1,\oa_2)$ is the Weierstrass P-function, 
$2\oa_{1,2}$ being a pair of periods, and
$\lambda$ is the (relativistic) deformation parameter.   
As shown by Ruijsenaars and Schneider in 
\cite{Ruijs1,Ruijs2}, this multi-particle 
model is integrable, and carries a representation of the 
Poincar\'e algebra in two dimensions. Moreover, a large number 
of the characteristics of the CM model are generalized in a 
natural way to the relativistic case, such as the existence 
of a Lax pair, a sufficient number of integrals of the motion 
in involution, and exact solution schemes in the special cases of 
rational and trigonometric/hyperbolic limits. The elliptic case 
has recently been investigated by Krichever and Zabrodin in \cite{KZ} 
in connection with the non-abelian Toda chain. 

In \cite{NRK} there was 
constructed an exact time-discretization of the equations
(\ref{eq:RCMa}) given by an integrable correspondence 
of the form 
\be \label{eq:dRS} 
\prod_{k=1 \atop k\ne \ell}^N \frac{ \sg(q_\ell-q_k+\ld)}{ \sg(
q_\ell-q_k-\ld)} = \prod_{k=1}^N \;\frac{\sg(q_\ell-\widetilde{q}_k)
\;\sg(q_\ell-\undertilde{q_k}+\ld)}{
\sg(q_\ell-\undertilde{q_k})\;\sg(q_\ell-\widetilde{q}_k-\ld)}
\  \ ,\  \ \ell=1,\dots,N\  .  
\ee 
In (\ref{eq:dRS}) the $q_k$ denote the particle positions for the time 
variable equal to $n$, the tilde being 
a shorthand notation for the discrete-time shift, i.e. for $q_k(n)=q_k$ 
we write $q_k(n+1)=\widetilde{q}_k$, and $q_k(n-1)=\undertilde{q_k}$. 
The function $\sg(x)$ is the Weierstrass 
sigma-function, (see Appendix for the definition), 
and $\ld$ is the parameter of the system as in the 
continuous case (\ref{eq:RCM}). 

The initial value problem for eqs. (\ref{eq:dRS}), given initial particle 
positions 
$\{ q_i(0)\}$ and $\{ q_i(1)\}$, leads to the problem of solving at each 
iteration step 
a coupled system of $N$ algebraic equations 
for $N$ unknowns, and it was shown in \cite{NRK} that in fact it is 
an integrable  symplectic correspondence (for a definition, see e.g. 
\cite{Ves}) 
with respect to the standard symplectic form $\ \Omega=\sum_k 
dp_k \wedge dq_k\ $. This 
implies that any branch of the correspondence given by 
eqs. (\ref{eq:dRS}) defines a 
canonical 
transformation with respect to the standard Poisson brackets given by
\be\label{eq:pb} 
\{ p_k,q_\ell \} = \dd_{k\ell}\   \ ,\   \ \{ p_k,p_\ell \} = 
\{ q_k,q_\ell \} = 0\  . 
\ee 
Here 
\be\label{eq:mom} 
p_\ell = \sum_{k=1}^{N} \left( -\log | \sg(q_\ell - \undertilde{q_k}) | + 
\log | \sg(q_\ell - \undertilde{q_k} + \ld) |  \right)\  .  
\ee
The discrete equations of motion (\ref{eq:dRS}) arise from a 
discrete Lax pair of the form
\bse \label{eq:lax} \bea 
L_\kp &=& \sum_{i,j=1}^N h_i^2 \Phi_\kp(q_i - q_j + \ld) e_{ij}\   , 
\label{eq:laxa} \\
M_\kp &=& \sum_{i,j} \widetilde{h}_i^2 \Phi_\kp(\widetilde{q}_i 
- q_j + \ld) e_{ij}\   , \label{eq:laxb} 
\eea \ese 
using the discrete Lax equation
\be \label{eq:Laxeq}
\widetilde{L}_\kp M_\kp = M_\kp L_\kp\   . 
\ee 
Notice here that in (\ref{eq:lax}) we use a different gauge
from the symmetric one used in \cite{NRK}. 
In eqs. (\ref{eq:lax}) the variable $\kp$ is an additional spectral 
parameter, and 
the matrices $e_{ij}$ are the standard elementary matrices whose 
entries are given by $(e_{ij})_{k\ell}=\dd_{ik}\dd_{j\ell}$. 
The function $\Phi_\kp$ is called the Baker function and is defined as
\be \label{eq:Phi} \Phi_\kp(x)\equiv \frac{\sg(x+\kp)}{\sg(x)\sg(\kp)}\  , 
\ee 
which obeys a number of functional relations listed in the Appendix. 
The auxiliary variables $h_\ell^2$ can be expressed in terms of the 
canonical variables, we obtain 
\be \label{eq:hp} 
h_\ell^2 = e^{p_\ell} \prod_{k\ne \ell} \frac{\sg(q_\ell - q_k -\ld)}{ 
\sg(q_\ell-q_k)} \  . 
\ee
In terms of these variables we have the following Poisson brackets:
\bea 
&&\{ q_k,q_\ell\} = 0\  \ ,\  \ \{ \log h_k^2\,, q_\ell\} = \dd_{k\ell}\  , 
\nn \\ 
&&\{ \log h_k^2\,,\,\log h_\ell^2\} = \zeta( q_k - q_\ell + \ld) 
+ \zeta( q_k - q_\ell - \ld) - 2 \zeta(q_k - q_\ell)\,,\qquad
k\neq\ell\,   . \label{eq:qhpb} 
\eea
It is easy to see that in terms of the canonical variables $p_\ell$ and 
$q_\ell$, the Lax matrix $L_\kp$ in (\ref{eq:laxa}) is exactly the same as 
the one of the continuous RS model, cf. \cite{BC}.  In fact, taking the 
continuum limit 
on the discrete-time part of the Lax pair (\ref{eq:lax}), namely 
the matrix $M_\kp$ (\ref{eq:laxb}), 
we obtain a Lax pair for the continuous RS model 
given by equations (\ref{eq:RCMa}). 
Since the $L_\kappa$-matrix 
for the discrete and continuous models is the same, 
the proof of involutivity of the invariants (integrals) ~$I_\ell = 
{\rm tr} L_\kappa^\ell$~ is the same in both cases, and sufficient to assess 
the Liouville integrability both discrete as well as continuous. 
The proof can be found in the original paper of Ruijsenaars, \cite{Ruijs2}, 
but is rather involved. Having at one's disposal an $r$-matrix structure 
would make the involutivity manifest. So far such an $r$-matrix has not 
been found in the full elliptic case. We will proceed now to 
establish this $r$-matrix structure.

\section{Classical $r$-matrix structure} 
\setcounter{equation}{0} 

As was noted recently by Suris, cf. \cite{Sur}, the main difference between 
the $r$-matrix structures of the relativistic and non-relativistic CM 
models resides in the fact that the latter is given in terms of a linear 
Lie-Poisson structure/bracket, 
whereas the former is given in terms of a quadratic bracket, 
cf. also \cite{BB}. The Poisson structure for the RS model will thus be 
given in the following quadratic $r$-matrix form (cf. \cite{Maillet,FM})
\bea \label{eq:PBs} 
\{\,L_\kp\,\stackrel{\otimes}{,}\,L_{\kp^\prime}\,\} &=& 
L_\kp \otimes L_{\kp^\prime} r_{\kp,\kp^\prime}^-  -  
r_{\kp,\kp^\prime}^+ L_\kp \otimes L_{\kp^\prime}  \nn \\ 
&&  + \left( L_\kp \otimes {\bf 1} \right) s_{\kp,\kp^\prime}^+  
\left( {\bf 1}\otimes L_{\kp^\prime} \right) -  
\left( {\bf 1}\otimes L_{\kp^\prime} \right) s_{\kp,\kp^\prime}^-  
\left( L_\kp \otimes {\bf 1} \right)\   .  
\eea 
The following symmetry conditions must hold for the $r$-matrices: 
$r_{\kp,\kp^\prime}^\pm$ and $s_{\kp,\kp^\prime}^\pm$
\bse \label{eq:rsconds}\be \label{eq:rscond1} 
P r_{\kp,\kp^\prime}^\pm P = - r_{\kp^\prime,\kp}^\pm \,,\qquad  
P s_{\kp,\kp^\prime}^+ P = s_{\kp^\prime,\kp}^-\   , 
\ee 
where $P$ is the permutation matrix in the tensor product of two
matrices, i.e. 
$P(A\otimes B)P=B\otimes A$, 
as well as the condition 
\be \label{eq:rscond2} 
r_{\kp,\kp^\prime}^+ - s_{\kp,\kp^\prime}^+ = 
r_{\kp,\kp^\prime}^- - s_{\kp,\kp^\prime}^-\  , 
\ee \ese 
in order that the quadratic Poisson algebra generates Hamiltonian flows for 
the 
invariants of the model, cf. \cite{Sur,NCP}. 
The condition (\ref{eq:rscond2}) was also 
formulated in \cite{NCP} in order to allow for a 
quadratic algebra on the lattice in terms 
of a local Lax representation to be integrated to a quadratic algebra 
in terms of the monodromy matrix. 

The choice of a gauge for the Lax matrices seems to be quite important 
in that it influences to a great extent the complexity of the 
associated $r$-matrix.  
The Lax matrix (\ref{eq:laxa}) has the nice property that 
it yields a remarkably simple  
$r$-matrix structure even in the elliptic case. 
In fact, the $r$-matrices we found are of the form
\bse \label{eq:rs} 
\be 
r_{\kp,\kp^\prime}^- = r_{\kp,\kp^\prime} -s_{\kp} +
Ps_{\kp^\prime}P\,, 
\ee
\be
r_{\kp,\kp^\prime}^+ = r_{\kp,\kp^\prime} +u^++u^-\,, 
\ee
\be
s_{\kp,\kp^\prime}^+ = s_{\kp}+u^+\,, 
\ee
\be
s_{\kp,\kp^\prime}^- = Ps_{\kp^\prime}P -u^-\,,
\ee \ese 
where \footnote{ 
In (\ref{eq:s0}) we mean by the  matrix 
~$\partial_\ld L_\kp$~ the following 
\[  
\partial_\ld L_\kp = \sum_{i,j=1}^N h_i^2 \Phi_\kp(q_i-q_j+\ld ) 
\left[ \zeta(\kp +q_i-q_j +\ld) - \zeta(q_i-q_j+\ld)\right]e_{ij} \   , 
\]   
i.e. we differentiate only with respect to the explicit dependence 
on the parameter $\ld$.} 
\bse\label{eq:rstu}\bea 
r_{\kp,\kp^\prime}&=&r_{\kp,\kp^\prime}^0+
\sum_{i} \zeta(\kp -\kp^\prime) e_{ii}\otimes e_{ii}
+ \sum_{i\neq j} \zeta(q_i -q_j) e_{ii}\otimes e_{jj}\,,\label{xx}\\
r_{\kp,\kp^\prime}^0 &=& \sum_{i\neq j} \Phi_{\kp -\kp^\prime}(q_i -q_j) 
e_{ij}\otimes e_{ji}\   ,  \label{eq:r0} \\ 
s_\kp  &=& \sum_{i,j} \left( L_\kp^{-1}\partial_\ld L_\kp \right)_{ij}  
e_{ij}\otimes e_{jj}\   ,  \label{eq:s0} \\ 
u^\pm  &=& \sum_{i,j} \zeta(q_j-q_i\pm\ld) 
e_{ii}\otimes e_{jj}\   .  \label{eq:u}  
\eea\ese 
The matrix elements in (\ref{eq:s0}) can be calculated 
explicitly
using the formula (\ref{eq:ACMinv}) for the inverse of the elliptic 
Cauchy matrix as well as making diligent use of the elliptic 
Lagrange interpolation formulae (\ref{eq:Lagr}) and (\ref{eq:Lagr2}), 
and this yields the following expression
\bea
&&\left( L_\kp^{-1}\partial_\ld L_\kp \right)_{ij} = 
\dd_{ij}\left[ \zeta(\kp+N\ld) -\zeta(\ld) + \sum_{k\neq i} 
\left( \zeta(q_i-q_k-\ld) - \zeta(q_i-q_k)\right)   
\right] \qquad \qquad\nn \\  
&&\qquad \qquad + (1-\dd_{ij}) \left[ \prod_{k=1\atop k\neq i}^N 
\frac{\sg(q_i-q_k-\ld)}{\sg(q_i-q_k)}\right] 
\left[ \prod_{k=1\atop k\neq j}^N 
\frac{\sg(q_j-q_k)}{\sg(q_j-q_k-\ld)}\right] \Phi_{\kp+N\ld}(q_i-q_j)\ . 
\label{eq:LdLa} 
\eea
The proof of the $r$-matrix structure (\ref{eq:PBs}) together with 
(\ref{eq:rs}) and 
(\ref{eq:rstu}) is by direct computation starting from the explicit form
of the $L$-matrix (\ref{eq:laxa}) and the Poisson brackets (\ref{eq:qhpb})
and making use of a number of elliptic relations which 
are listed in the Appendix. We will not give any details, but just 
restrict ourselves to giving a few intermediate relations, 
which can be established using the formulas from the Appendix, namely:
\bse\label{eq:rels} \bea 
\left( L_\kp\otimes L_{\kp^\prime}\right) s_\kp &=& 
\sum_{ij}\sum_{\ip\jp} h_i^2h_\ip^2 \Phi_\kp(q_i-q_j+\ld) 
\Phi_{\kp^\prime}(q_\ip-q_\jp+\ld)\,e_{ij}\otimes e_{\ip\jp}\nn\\ 
&\times &
\dd_{j\jp} \left[ \zeta(\kp +q_i-q_j +\ld) - \zeta(q_i-q_j+\ld)\right] 
\   ,  \label{eq:relsa} \\ 
\left( L_\kp\otimes {\bf 1}\right) s_\kp 
\left( {\bf 1}\otimes L_{\kp^\prime}\right) &=& 
\sum_{ij}\sum_{\ip\jp} h_i^2h_\ip^2 \Phi_\kp(q_i-q_j+\ld) 
\Phi_{\kp^\prime}(q_\ip-q_\jp+\ld)\,e_{ij}\otimes e_{\ip\jp}\nn\\ 
&\times & 
\dd_{j\ip} \left[ \zeta(\kp +q_i-q_j +\ld) - \zeta(q_i-q_j+\ld)\right] 
\   ,  \label{eq:relsc} 
\eea 
as well as 
\bea 
&& \left[\,r_{\kp,\kp^\prime}^0\,,\, L_\kp\otimes L_{\kp^\prime}\right] = 
\sum_{ij}\sum_{\ip\jp} h_i^2h_\ip^2 \Phi_\kp(q_i-q_j+\ld) 
\Phi_{\kp^\prime}(q_\ip-q_\jp+\ld)\,e_{ij}\otimes e_{\ip\jp} \nn \\ 
&&~~~ \times \left\{ \,(1-\dd_{i\ip})(1-\dd_{j\jp}) \left[ 
\zeta(q_i-q_\ip) + \zeta(q_\ip-q_j+\ld) + \zeta(q_j-q_\jp) 
- \zeta(q_i-q_\jp+\ld)\right] \right. \nn \\ 
&& ~~~ + \dd_{j\jp}(1-\dd_{i\ip})\left[  
\zeta(q_i-q_\ip) - \zeta(\kp+q_i-q_j+\ld) + 
\zeta(\kp^\prime + q_\ip-q_\jp+\ld ) + \zeta(\kp-\kp^\prime)\right] \nn \\ 
&& ~~~ \left. + \dd_{i\ip}(1-\dd_{j\jp})\left[  
\zeta(q_j-q_\jp) + \zeta(\kp+q_i-q_j+\ld) - 
\zeta(\kp^\prime + q_\ip-q_\jp+\ld ) - \zeta(\kp-\kp^\prime)\right] 
\,\right\}\,. \nn \\ 
 \label{eq:relsd} 
\eea \ese
Remark here that our $r$-matrices do not depend on momenta,
like in the non-relativistic case \cite{Skly}, which was 
the motivation for the choice of the gauge of $L_\kappa$.

As a direct application of the $r$-matrix structure let us calculate 
the (continuous) time part of the Lax representation. It is obtained from the 
following formula:
\be
({\rm tr}\otimes id)(L_\kappa\otimes {\bf 1})(r_{\kappa,\kappa^\prime}^+- 
s_{\kappa,\kappa^\prime}^+)=\Phi_{\kappa-\kappa^\prime}(\lambda)
L_{\kappa^\prime}-\Phi_{\kappa}(\lambda)N_{\kappa^\prime}
\ee 
where
\be
N_\kappa=\sum_i\left[\zeta(\kappa)h_i^2+\sum_{j\neq i}h_j^2
\zeta(q_i-q_j)-\sum_jh_j^2\zeta(q_i-q_j-\ld)\right]e_{ii}
+\sum_{i\neq j}h_i^2\Phi_\kappa(q_i-q_j)e_{ij}\,,
\ee
which, together with (\ref{eq:laxa}), leads to the Lax representation 
found in \cite{BC} 
for the continuous RS model (up to a gauge transformation!).  
Thus, the continuous equations of motion 
(\ref{eq:RCMa}) corresponding to the Hamiltonian ${\rm tr}\,L_{\kappa}$
follow from the Lax equation
\be \dot{L}_\kp = \left[ N_{\kp}\,,\,L_\kp \right]\,. 
\ee 

\paragraph{Remarks:}
\begin{itemize}
\item The non-relativistic limit is obtained by letting ~$\ld\rightarrow 0$~ 
while scaling the momenta $p_i:=\lambda p_i$ and making the canonical
transformation $p_i:=p_i+\sum_{k\neq i}\zeta(q_i-q_k)$
such that ~$h_i^2\rightarrow 1+\lambda p_i+O(\lambda^2)$~ 
in (\ref{eq:hp}). The $r$-matrix structure is linear in that limit since the 
$L$-matrix behaves as 
\[ L_\kp\  \ \rightarrow\  \ \ld^{-1}+\zeta(\kappa) + \sum_i p_i e_{ii} 
+ \sum_{i\neq j} \Phi_\kp(q_i-q_j) e_{ij} + O(\ld)\  ,  \] 
whereas  the matrices $r_{\kp,\kp^\prime}^\pm\,,\,s_{\kp,\kp^\prime}^\pm$ 
enter in the following combination 
\be
r_{\kappa,\kappa^\prime}^+
+s_{\kappa,\kappa^\prime}^-\ \ \rightarrow\ \ 
r_{\kp,\kp^\prime}^{\rm (nr)} + O(\ld)\   , 
\ee 
in which the non-relativistic $r$-matrix is given by
\bea 
r_{\kp,\kp^\prime}^{\rm (nr)}& =&  
\sum_i\left(\zeta(\kappa-\kappa^\prime)+\zeta(\kappa^\prime)\right)
e_{ii}\otimes e_{ii}+\sum_{i\neq j}\Phi_{\kappa-\kappa^\prime}(q_i-q_j)
e_{ij}\otimes e_{ji} \nn \\ 
&+& \sum_{i\neq j}\Phi_{\kappa^\prime}(q_i-q_j) e_{jj}\otimes e_{ij}\    , 
\eea  
thus recovering the result of \cite{Skly} in the leading terms.
\item It does not seem easy to compare our result with that of 
Avan and Rollet in \cite{AR} 
because of the not so transparent nature of that result (they wrote 
their $r$-matrix structure in a linear form hiding the quadratic 
nature of the RS model). In the trigonometric limit taking the special 
parameter value ~$e^{2\ld}=-1$~ the structure we found should reduce to the 
one given earlier by Babelon and Bernard in \cite{BB}. 
\item In \cite{Sur} the author seems to suggest that the $r$-matrix 
for the relativistic model and the non-relativistic model is the same 
in the rational and trigonometric/hyperbolic limits, 
which is demonstrated by an independence of the $r$-matrix objects 
on the relativistic parameter $\ld$. This, however, seems to be 
no longer true in the case of the elliptic model, and there is no apparent 
way in which one can get rid of the $\ld$-dependence in the explicit 
formulas (\ref{eq:rs}) and (\ref{eq:rstu}). 
\item We do not write down any Yang-Baxter type relations between the 
$r$-matrices given in (\ref{eq:rs}), because as a consequence of the 
dynamical nature the Yang-Baxter algebra does not seem to be closed, i.e. 
the Yang-Baxter 2-cocycle consists of terms which contain Poisson brackets 
with the $L$-matrix itself. It is an interesting open problem to see whether 
one can close the algebra on any level, leading to a possible truncation 
of some higher-order Yang-Baxter cocycle. So far, no results along this 
direction exist. 
\end{itemize} 
 
\section{Conclusions and outlook}

In this paper we have presented the classical $r$-matrix structure of the full
elliptic Ru\-ij\-se\-naars-Schneider 
model. Since the model is the most general 
among the Calogero-Moser type models for the $sl_n$ Lie algebra, our result is
in a sense conclusive. Still, a number of questions wait to be answered.
Since the dynamical nature of the $r$-matrices implies that the corresponding
Yang-Baxter algebra is not closed, it is not yet clear how to use it for 
quantization. 

Concerning the quantization  problem, 
the recent result of Hasegawa \cite{Hase} 
should be mentioned, who has found an 
interesting connection between the quantum $L$-operator
associated with Belavin's $R$-matrix and the quantum integrals of the 
Ruijsenaars' model. This somehow implies that on the classical level there
should exist a gauge transformation involving the dynamical variables between
an elliptic $r$-matrix of Belavin type and the one we have constructed in the
present paper. 

Another result concerns the separation of variables approach
leading to the explicit integral representations for the Macdonald polynomials 
associated with the trigonometric RS model, cf.\ \cite{KS}. So far 
the only result for the elliptic case is the separation of variables
for the 3-particle nonrelativistic Calogero-Moser model \cite{Skl95}.
The elliptic $r$-matrix could presumably help in constructing a separation
of variables in the general case.

One more possible application of the results of this paper could 
lie in the time discretization of the RS model constructed 
in \cite{NRK}. 
One feature of the proposed time discretization 
is that these discrete models share the 
time-independent 
part of the Lax pair with the corresponding continuous models and, 
consequently, the 
invariants take the same form in both cases. Thus, the proof of the 
Liouville integrability  
(or the involutivity of the invariants) is exactly the same as for the 
continuous model. For the discrete models it 
seems that the most prominent role is played by the 
$M$-matrix. In fact, in similar systems related to integrable 
lattices it was found that there exists an extended Yang-Baxter structure 
which incorporates the $M$-matrix as well as the $L$-matrix in the 
Yang-Baxter algebra. 
Now that a classical $r$-matrix structure is available for the generic 
RS system, it would  be interesting to search for similar extended 
YB structures for these many-body systems as well. This would possibly 
yield some new insights into the problem of performing the $R$-matrix 
quantisation for dynamical $r$-matrices as the one we have obtained 
in the present paper. 

\subsection*{Acknowledgements} VBK is supported by EPSRC under Grant No. GR/K63887. 
OR is grateful to the Department of Applied Mathematical Studies 
of the University of Leeds for its hospitality during his visit. 

\subsection*{Appendix:  Formulas for elliptic functions} 
\def\theequation{A.\arabic{equation}}
\setcounter{equation}{0}

Here, we collect some useful formulas for elliptic functions, see also 
the standard textbooks e.g. \cite{WW}. 
The Weierstrass sigma-function is defined by 
\be
\sigma(x) = x \prod_{(k,\ell) \ne (0,0)} \left(1-\frac{x}{\omega_{k\ell}}
\right)
 \exp\left[ \frac{x}{\omega_{k\ell}} + \frac{1}{2} 
( \frac{x}{\omega_{k\ell}})^2\right]\ , 
\ee
with $\oa_{kl}=2k\oa_1 + 2\ell \oa_2$ and  
$2\omega_{1,2}$  being a fixed pair of the primitive periods.  
The relations between the Weierstrass elliptic functions are given by 
\be 
\zeta(x) =  \frac{\sigma^\prime(x)}{\sigma(x)}\   \  , \   \ 
\wp(x) = - \zeta^\prime(x)\   ,  \ee 
where $\sg(x)$ and $\zeta(x)$ are odd functions and $\wp(x)$ is an 
even function of its argument. 
We recall also that the $\sg(x)$ is an entire function, and 
$\zeta(x)$ is a meromorphic function having simple poles at 
$\omega_{kl}$, both being quasi-periodic, obeying  
\[ 
\zeta(x+2\omega_{1,2}) = \zeta(x) + 2\eta_{1,2}\    \ ,\    \ 
\sigma(x+2\omega_{1,2}) = -\sigma(x)
e^{2\eta_{1,2}(x+\omega_{1,2})}\  ,
\] 
in which $\eta_{1,2}$ satisfy ~$\eta_1\omega_2 - \eta_2\omega_1 
= \frac{\pi i}{2}$~, whereas $\wp(x)$ is doubly periodic.
{}From an algebraic point of view, the most important property of 
these elliptic functions is the existence of a number of functional 
relations, the most fundamental being 
\be \label{eq:zs} 
\zeta(\ar) + \zeta(\bb) + \zeta(\gm) - \zeta(\ar +\bb + \gm)
  = \frac{  \sigma(\ar + \bb )  \sigma(\bb + \gm ) 
\sigma( \gm + \ar) }{ \sigma(\ar) \sigma(\bb) \sigma(\gm) 
\sigma(\ar + \bb + \gm )}~ , 
\ee
which can also be cast into the following form
\be\label{eq:12} 
\Phi_\kp(x)\Phi_\kp(y) = 
\Phi_{\kappa}(x+y)\left[ \zeta(\kp) +\zeta(x) +\zeta(y) -\zeta(\kp +x+y)
\right] \   ,  
\ee
The following three-term relation for $\sigma(x)$ is a consequence of 
(\ref{eq:zs}) 
\bea
\sigma(x+y) \sigma(x-y) \sigma(a+b) \sigma(a-b) &=& \sigma(x+a) \sigma(x-a) 
\sigma(y+b) \sigma(y-b)  \nn  \\ 
&& ~ - \sigma(x+b) \sigma(x-b) \sigma(y+a) \sigma(y-a)\   ,  \label{eq:8} 
\eea 
and this equation can be cast into the following convenient form 
\be\label{eq:14} 
\Phi_\kappa(x)\Phi_\ld(y) = 
\Phi_{\kappa}(x-y)\Phi_{\kappa+\ld}(y) + \Phi_{\kappa +\ld}(x)
\Phi_{\ld}(y-x)\   ,  
\ee
which is obtained from the elliptic analogue of the partial 
fraction expansion, i.e. eq. (\ref{eq:12}). 

There are few additional 
important identities that are used in the proof of the 
$r$-matrix structure, the main one is given by 
\bea  
&&\Phi_{\kp-\kp^\prime}(a-b)\Phi_\kp(x+b)\Phi_{\kp^\prime}(a+y) 
- \Phi_{\kp-\kp^\prime}(x-y)\Phi_\kp(y+a)\Phi_{\kp^\prime}(x+b) =  \nn \\ 
&& ~ = \Phi_\kp(x+a)\Phi_{\kp^\prime}(y+b) \left[ 
\zeta(a-b) + \zeta(x+b) - \zeta(x-y) - \zeta(y+a) \right] \  ,\label{eq:15}  
\eea  
which can be derived from (\ref{eq:14}) together with (\ref{eq:12}), and  
\bea  
&&\Phi_{\kp-\kp^\prime}(x-y)\Phi_\kp(y+a)\Phi_{\kp^\prime}(x+a) = 
\qquad\qquad\nn\\ 
&& = \Phi_\kp(x+a)\Phi_{\kp^\prime}(y+a) \left[ 
\zeta(x-y) - \zeta(\kp+x+a) + \zeta(\kp^\prime+y+a) + 
\zeta(\kp-\kp^\prime) \right] .  \label{eq:16}
\eea 
It is eqs. (\ref{eq:15}) and (\ref{eq:16}) that are used in the 
derivation of (\ref{eq:relsd}) which forms the main step 
in the computation of the $r$-matrix. 

In \cite{NRK} there was used an elliptic version of 
the Lagrange interpolation formula, which was derived on the basis
of an elliptic version of the Cauchy identity. We can write the elliptic
Cauchy identity in the following elegant form: 
\be \label{eq:cauchy} 
\det\left( \Phi_\kappa(x_i - y_j)\right) = 
\Phi_\kp( \Sigma ) \sg(\Sigma ) 
\frac{ \prod_{k<\ell} \sigma(x_k - x_\ell) 
\sigma(y_\ell - y_k) }{  
\prod_{k,\ell} \sigma(x_k - y_\ell) }\,,\    \ {\rm where}\  \ 
\Sigma\equiv 
\sum_i (x_i-y_i)\   . 
\ee 
An elliptic form of the Lagrange interpolation formula is obtained by 
expanding (\ref{eq:cauchy}) along one of its rows or columns. Thus, 
we obtain
\be 
\prod_{i=1}^N \frac{\sigma(\xi - x_i)}{\sigma(\xi - y_i)}\,=\,
\sum_{i=1}^N \Phi_{-\Sigma}(\xi - y_i) 
\frac{\textstyle \prod_{j=1}^N \sigma(y_i - x_j)}{\textstyle 
\prod_{j=1\atop j\ne i}^N \sigma(y_i - y_j)}\,,\    \ 
 {\rm when}\ \ \Sigma =\sum_{i=1}^N (x_i -y_i)\neq 0\,, 
\label{eq:Lagr} \ee  
and 
\be
\prod_{i=1}^N \frac{\sigma(\xi - x_i)}{\sigma(\xi - y_i)}\,=\,
\sum_{i=1}^N \left[ \zeta(\xi - y_i)  
- \zeta(x - y_i)\right]  
\frac{\textstyle \prod_{j=1}^N \sigma(y_i - x_j)}{\textstyle 
\prod_{j=1\atop j\ne i}^N \sigma(y_i - y_j)}\,,\quad
 {\rm when}\quad \sum_{i=1}^N (y_i -x_i) = 0\,, 
\label{eq:Lagr2} \ee
(here $x$ denotes one of the zeros $x_i$). 
Note that in this case the left hand side is a meromorphic 
function on the elliptic curve as a consequence of Abel's 
theorem. It can be easily verified that eq. 
(\ref{eq:Lagr2}) is independent of the choice of $x$ 
as a consequence of the relation
\be
\sum_{i=1}^N 
\frac{\textstyle \prod_{j=1}^N \sigma(y_i - x_j)}{\textstyle 
\prod_{j=1\atop j\ne i}^N \sigma(y_i - y_j)}\,=\,0\   \quad
{\rm when} \quad 
~~  \ \sum_{i=1}^N (y_i -x_i) = 0\,, 
\ee
(cf. e.g. \cite{WW}, p. 451), which follows from eq. 
(\ref{eq:Lagr}) in the limit $\Sigma\rightarrow 0$. 

Finally, we give the expression for the inverse of the elliptic 
Cauchy matrix, namely 
\be \label{eq:ACMinv}
\left[ \left( \Phi_\kp(x_\cdot - y_\cdot ) \right)^{-1}\right]_{ij} = 
\Phi_{\kp+\Sigma}(y_i-x_j) \frac{ P(y_i) Q(x_j)}{Q_1(y_i)P_1(x_j)}\   , 
\ee  
(with $\Sigma$ as before), 
in terms of the elliptic polynomials  
$$ P(\xi) = \prod_{k=1}^N \sg(\xi -x_k) \     \ ,\     \ 
Q(\xi) = \prod_{k=1}^N \sg(\xi -y_k)\    , $$  
and
\be
P_1(x_j)=\prod_{k\neq j}\sigma(x_j-x_k)\,, \qquad
Q_1(y_i)=\prod_{k\neq i}\sigma(y_i-y_k)\,. 
\ee
Equation (\ref{eq:ACMinv}) can be derived using (\ref{eq:Lagr})
and (\ref{eq:Lagr2}), and it is used to derive equation
(\ref{eq:LdLa}) in the main text. 

\pagebreak


\begin{thebibliography}{99}
\bibitem{AT}
J. Avan and M. Talon, {\em Classical $R$-matrix structure for the Calogero 
model}, Phys. Lett. {\bf B303} (1993) 33--37. 
\bibitem{Skly} 
E.K. Sklyanin, {\em Dynamical $r$-matrices for elliptic Calogero-Moser 
model}, 
St. Petersburg J. Math. {\bf 6} (1994) 227. 
\bibitem{BS}
H.W. Braden and T. Suzuki, {\em $R$-matrices for the elliptic Calogero-Moser 
models}, Lett. Math. Phys. {\bf 30} (1994) 147. 
\bibitem{EEKT}
J.C. Eilbeck, V.Z. Enol'skii, V.B. Kuznetsov and A.V. Tsiganov,
{\em Linear $r$-matrix algebra for classical separable systems},
J. Phys. A: Math.Gen. {\bf 27} (1994) 567. 
\bibitem{ABB}
 J.~Avan, O.~Babelon and E.~Billey, {\it The Gervais-Neveu-Felder equation 
and the quantum Calogero-Moser systems}, preprint PAR LPTHE 95-25, May 1995;
{\sf hep-th/9505091} (to be published in Commun.\ Math.\ Phys.)
\bibitem{Ruijs1} 
S.N.M. Ruijsenaars and H. Schneider, 
{\em A new class of integrable systems and its relation to solitons},
Ann. Phys. {\bf 170} (1986) 370--405. 
\bibitem{Ruijs2}
S.N.M. Ruijsenaars, {\em Complete integrability of relativistic 
Calogero-Moser systems and elliptic function identities}, 
Comm. Math. Phys. {\bf 110} (1987) 191--213. 
\bibitem{Diej}
J.-F. van Diejen, {\em Commuting difference operators with polynomial 
eigenfunctions}, Compos. Math. {\bf 95} (1995) 183--233. 
\bibitem{Macd}
I.G. Macdonald,
{\em Symmetric functions and Hall polynomials}, 2nd Ed., (Oxford Univ. Press, 
1995). 
\bibitem{AR}
J. Avan and G. Rollet, {\em The classical $r$-matrix for the relativistic 
Ruijsenaars-Schneider system}, Phys. Lett. {\bf A212} (1996) 50--54. 
\bibitem{BB}
O. Babelon and D. Bernard, {\em The sine-Gordon solitons as 
a N-body problem}, Phys. Lett. {\bf B317} (1993) 363--368. 
\bibitem{Sur}
Yu.B. Suris, {\em Why are the Ruijsenaars-Schneider and the Calogero-Moser 
hierarchies governed by the same $r$-matrix?}, Preprint Univ. of Bremen, 
February 1996, {\tt hep-th/9602160}. 
\bibitem{KZ}
I. Krichever and A. Zabrodin, {\em Spin generalization of the 
Ruijsenaars-Schneider 
model, non-Abelian 2D Toda chain and representations of Sklyanin algebra}, 
Preprint Landau Institute, {\tt hep-th/9505039}. 
\bibitem{NRK}
F.W. Nijhoff, O. Ragnisco and V.B. Kuznetsov,
{\em Integrable time-discretization of the Ruijsenaars-Schneider model},
Commun. Math. Phys. {\bf 176} (1996) 681--700.  
\bibitem{Ves}
A.P. Veselov, {\em Growth and integrability in the dynamics of maps}, 
Commun. Math. Phys. {\bf 145} (1992) 181--193.
\bibitem{BC}
M. Bruschi and F. Calogero, {\em The Lax representation for an integrable 
class of relativistic dynamical systems}, Commun. Math. Phys. {\bf 109} 
(1987) 481--492. 
\bibitem{Maillet}
J.-M.~Maillet, {\it Kac-Moody algebra and extended Yang-Baxter relations 
in the $O(N)$ non-linear $\sigma$-model}, 
Phys.\ Lett.\ {\bf B162} (1985) 137--142.
\bibitem{FM}
L.~Freidel and J.-M.~Maillet, {\it Quadratic algebras and integrable systems},
Phys.\ Lett.\ {\bf B262} (1991) 278--284.
\bibitem{NCP}
F.W. Nijhoff, V.G. Papageorgiou and H.W. Capel, in {\em Quantum groups}, 
ed. P.P.~Ku\-li\-sh, Springer Lect. Notes Math. {\bf 1510} (1992) 312.
\bibitem{Hase}
K. Hasegawa, {\em Ruijsenaars' commuting difference operators as commuting 
transfer matrices}, Preprint Tohoku University, 1995; {\tt q-alg/9512029.}
\bibitem{KS}
V.B. Kuznetsov and E.K. Sklyanin, 
{\em Separation of variables for $A_2$ Ruijsenaars model and 
new integral representation for $A_2$ Macdonald polynomials},
(Submitted to J.Phys.A), {\tt q-alg/9602023}. 
\bibitem{Skl95} 
 E.~K.~Sklyanin. {\it Separation of variables. New trends,}
Progr.\ Theor.\ Phys.\ Suppl.\ {\bf 118} (1995) 35--60.
\bibitem{WW} 
E.T. Whittaker and G.N. Watson, {\em A course in modern analysis}, 
(Cambridge University Press, 4th ed., 1988). 
\end{thebibliography}
\end{document}